\documentclass[12pt]{iopart}

\usepackage{graphicx}
\usepackage[numbers,sort&compress]{natbib}

\begin{document}
\title[Temporal and spatial superbunching effects]{Temporal and spatial superbunching effects from a pair of modulated distinguishable classical light}

\author{Sheng Luo$^1$, Huai-Bin Zheng$^{1,*}$,
Wan-Ting Xu$^2$, Yu-Chen He$^1$ Shuang-Hao Zhang$^1$ and Zhuo Xu$^1$}

\address{$^1$ Key Laboratory of Multifunctional Materials and Structures, Ministry of Education, School of Electronic Science and Engineering, Xi'an Jiaotong University, Xi'an, Shaanxi 710049, China}
\address{$^2$ School of Science, Xi'an Polytechnic University, Xi'an, Shaanxi 710048, China}
\address{$^*$ Author to whom any correspondence should be addressed}
\ead{huaibinzheng@xjtu.edu.cn}

\vspace{2pt}

\begin{indented}
\item[]October 2020
\end{indented}

\begin{abstract}
From the Feynman path integration theory of view, the Hanbury Brown-Twiss effect would not be observed for one definite two-photon propagation path, as well as the superbunching effect. Here, temporal and spatial superbunching effects are measured from a pair of modulated distinguishable classical light. These interesting phenomena are realized by passing two orthogonal polarized laser beams through two rotating ground glass plates in sequence. To understand the underlying physical process, the intensity fluctuation correlation theory is developed to describe the superbunching effect in the temporal and spatial domain, which agrees with experimental results well. Such experimental results are conducive to the study of superbunching effect which plays an important role in improving the performance in related applications, such as the contrast of ghost imaging.
\end{abstract}

%
%
%
%
%

\section{Introduction}

Since Hanbury Brown and Twiss (HBT) first observed the two-photon bunching effect in 1956 \cite{brown1956correlation,brown1956test}, they found that photons emitted by thermal light source are more inclined to arrive in pairs rather than randomly. The degree of second-order correlation of thermal light in the HBT interferometer was measured as 2, which greatly promoted the development of quantum optics \cite{glauber1963quantum,glauber1963coherent,glauber1963photon,sudarshan1963equivalence,glauber2006nobel}. Then the discovery of the superbunching effect that the degree of second-order correlation is greater than 2 has also attracted the attention of researchers \cite{lipeles1965direct,kaul1966observation,akhmanov1967nonlinear,kocher1967polarization,auffeves2011few,hoi2012generation,grujic2013repulsively,bhatti2015superbunching,hong2012two,zhou2017superbunching,manceau2019indefinite,Yu2019Experimental}. It is generally accepted that the HBT effect is all well explained by both classical and quantum interpretations which can be unified. Classical intensity fluctuation correlation theory was usually employed to interpret this phenomenon \cite{brown1957interferometry}, which emphasized whether the two detectors could detect specks with the same intensity fluctuation in the HBT interferometer. It can also be interpreted by quantum theory, in which two-photon bunching is interpreted by superposition of different alternatives to trigger a two-photon coincidence count \cite{glauber1963quantum,glauber1963coherent,glauber1963photon,fano1961quantum}. In 2017, Bin Bai et al. reported an experiment that bunching effect was observed without two-photon interference since there is only one alternative to trigger a joint detection event \cite{bai2017hanbury}. It seems to indicate the discrepancy between classical and quantum theory. Ultimately, he used Glauber's quantum optical correlation theory to explain this interesting phenomenon. Later, Yu Zhou et al. proposed quantum two-photon interference theory to explain superbunching pseudothermal light based on multi-path interference \cite{zhou2017superbunching}. However, there is no classical intensity fluctuation correlation theory analyzing the such superbunching effect with distinguishable interference paths in the temporal and spatial domain.

In this paper, we proposed and demonstrated a method to observe temporal and spatial superbunching effect from a pair of modulated distinguishable classical light. The normalized temporal and spatial second-order correlation function were measured as ${g^{\left(2\right)}}\left(0\right)=3.83\pm0.02$ and ${g^{\left(2\right)}}\left(0\right)=3.42\pm0.02$, respectively. At the same time, the
theoretical results derived from classical intensity fluctuation correlation theory are in good agreement with the experimental results. Based on above study, it further demonstrates the unity and complementarity between quantum theory and classical theory, which is not only replenish the theoretical study of superbunching effect, but also helpful to understand the physical essence of superbunching effect.

The remaining parts of the paper are organized as follows. In Sec. \ref{Methods}, classical intensity fluctuation correlation theory is employed to interpret the superbunching effect, then the theoretical derivation results is verified by the designed experimental scheme. The discussions and conclusions are in Sec. \ref{Discussion} and \ref{Conclusion}, respectively.

\section{\label{Methods}Methods}

\subsection{Classical intensity fluctuation correlation theory}
Instead of the quantum two-photon interference theory, the classical intensity fluctuation correlation theory is used to explain the superbunching effect. This is mainly due to the correlation between the electromagnetic waves propagating to two detectors located at different space-time coordinates $({r_1},{t_1})$ and $({r_2},{t_2})$. Therefore, the function of the second-order correlation of optical field can be expressed as
\begin{equation}\label{equ1}
{G^{(2)}}({r_1},{t_1};{r_2},{t_2}) = \left\langle {{E_1}({r_1},{t_1}){E_2}({r_2},{t_2}){E_1}^*({r_1},{t_1}){E_2}^*({r_2},{t_2})} \right\rangle,
\end{equation}
where ${\left\langle { \cdot  \cdot  \cdot } \right\rangle}$ is the ensemble average by taking all the possible realizations into account, ${E_i}({r_i},{t_i})$ is the electric field intensity on the surface of detector $\rm D_i$, where ${E_i}^*({r_i},{t_i})$is the complex conjugate of ${E_i}({r_i},{t_i})$, $i = 1,2$.

\begin{figure}[htb]
\centering
\includegraphics[width=10cm]{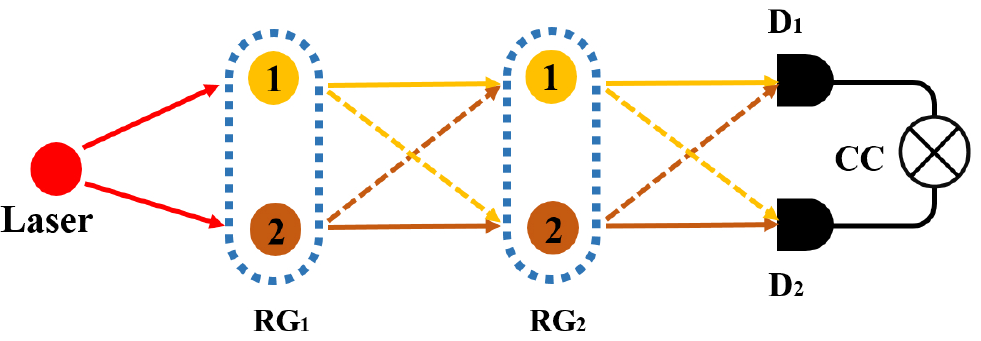}
\caption{\label{figure1} The theoretical model diagram of a joint detection event triggered by different electromagnetic waves. $\rm RG_1$ and $\rm RG_2$ are two rotating ground glass plates. $\rm D_1$ and $\rm D_2$ are two single-photon detectors in the HBT interferometer. $\rm CC$ is the two-photon coincidence count detection system.}
\end{figure}

When a single-mode continuous-wave laser light passes through two rotating ground glass (RG) plates continuously, the scattered pseudothermal light, which fluctuates in time and space, will arrive at the surfaces of the two detectors. The theoretical model of superbunching effect is shown in figure \ref{figure1}. There will be eight different electromagnetic waves triggering the two detectors, $E_{111}$, ${E_{112}}$, ${E_{121}}$, ${E_{122}}$, ${E_{211}}$, ${E_{212}}$, ${E_{221}}$, ${E_{222}}$. Taking $E_{121}$ as an example, it represents the electromagnetic wave scattered by the laser passing through position 1 of the $\rm RG_1$, then propagates to the position 2 of the $\rm RG_2$, finally arrives at the surface of $\rm D_1$. Other meaning of electromagnetic waves are analogized in turn. For the meaning of $E_{121}$, it is actually a compound electric field, which is obtained by multiplying a initial electromagnetic waves with the propagation function under different conditions. According to Green function for a point light source in classical optics \cite{Goodman1995Introduction}, the electromagnetic wave which arrived at the surface of detector without the spatial part can be expressed as
\begin{eqnarray}\label{equ2}
{E_{ijk}} & =E_{i} \cdot T_{j | i} \cdot T_{k|j|i} \nonumber \\
& =e^{-i\omega_{0}\left(t_{i}-t_{0}\right)} \int_{\omega_{0}-\frac{1}{2} \Delta \omega_{1}}^{\omega_{0}+\frac{1}{2} \Delta \omega_{1}} e^{-i\omega_{0j}\left(t_{j}^{''}-t_{i}^{'}\right)} d \omega_{0 j} \int_{\omega_{0}-\frac{1}{2}\Delta \omega_{2}}^{\omega_{0}+\frac{1}{2} \Delta \omega_{2}} e^{-i\omega_{k}\left(t_{k}-t_{j}^{''}\right)} d \omega_{k},
\end{eqnarray}
where $E_i$ is the electromagnetic wave which propagated by laser passing through position $i$ of the $\rm RG_1$, ${T_{j|i}}$ represents the propagation function of pseudothermal light when $E_i$ passes through the position $j$ of the $\rm RG_2$, and ${T_{k|j|i}}$ is the propagation function of electromagnetic wave when it arrives at the surface of $\rm D_k$ under the condition of ${E_i}\cdot{T_{j|i}}$. ${\omega _0}$ is the center frequency of the laser, ${\omega _{0i}}$ is the frequency which the laser scattered by the position $i$ of the $\rm RG_1$, and ${\omega _{j}}$ is the frequency of the light which scattered by the position $j$ of the $\rm RG_2$. $\Delta {\omega _{\rm{1}}}$ and $\Delta {\omega _{\rm{2}}}$ represent the spectral widths of pseudothermal light scattered by the $\rm RG_1$ and $\rm RG_2$, respectively. $t_i^{'}$, $t_j^{''}$ and $t_k$ are the time when the electromagnetic wave arrives at position $i$ of the $\rm RG_1$, position $j$ of the $\rm RG_2$ and the detector $k$ surface respectively, $i,j,k = 1,2$.

Then there are four types of coincidence current detected by joint detection system as shown in the figure \ref{figure1}, ${E_{111}}{E_{222}}$, ${E_{112}}{E_{221}}$, ${E_{121}}{E_{212}}$, ${E_{122}}{E_{211}}$. Therefore, the total current measured by the detector can be expressed as
\begin{equation}\label{equ3}
\fl {E_1}({r_1},{t_1}){E_2}({r_2},{t_2}) = \frac{1}{4}{E_{111}}{E_{222}} + \frac{1}{4}{E_{112}}{E_{221}} + \frac{1}{4}{E_{121}}{E_{212}} + \frac{1}{4}{E_{122}}{E_{211}}.
\end{equation}

Substituting equation (\ref{equ3}) into equation (\ref{equ1}), the second-order correlation function can be written as
\begin{eqnarray}\label{equ4}
\fl {G^{(2)}}({r_1},{t_1};{r_2},{t_2}) &= \frac{1}{{16}}\langle \left({{E_{111}}{E_{222}} + {E_{112}}{E_{221}}+{E_{121}}{E_{212}} + {E_{122}}{E_{211}}} \right)\nonumber\\&\cdot
\left({{E_{111}}^*{E_{222}}^* + {E_{112}}^*{E_{221}}^* + {E_{121}}^*{E_{212}}^* + {E_{122}}^*{E_{211}}^*} \right)\rangle.
\end{eqnarray}

There will be 16 terms after expanding equation (\ref{equ4}). It is obvious that there are 12 cross-correlation terms besides four auto-correlation terms. The whole second-order correlation function can be categorized into four groups. We calculate one term for each group, and the other terms are calculated in the same way. Substituting equation (\ref{equ2}) into equation (\ref{equ4}), one can get concrete expression.

The first group is all the autocorrelation group. The calculated results are all constant. Take $E_{111}E_{222}E_{111}^*E_{222}^*$ as a example
\begin{eqnarray}\label{equ5}
\fl \langle E_{111} E_{222} E_{111}^{*} E_{222}^{*}\rangle
& =\langle\left(E_{1} T_{1|1} T_{1|1|1}\right)\left(E_{2} T_{2 | 2} T_{2|2| 2}\right)\left(E_{1}^{*} T_{1 | 1}^{*} T_{1|1| 1}^{*}\right)\left(E_{2}^{*} T_{2 | 2}^{*} T_{2|2| 2}^{*}\right)\rangle \nonumber \\
& =\left(\Delta \omega_{1}\right)^{2}\left(\Delta \omega_{2}\right)^{2},
\end{eqnarray}
The rest of the three terms, ${E_{112}}{E_{221}}{E_{112}}^*{E_{221}}^*$,
${E_{121}}{E_{212}}{E_{121}}^*{E_{212}}^*$,
${E_{122}}{E_{211}}{E_{122}}^*{E_{211}}^*$ in the same group have the same result as the one of equation (\ref{equ5}).

The second term that needs to be calculated is $E_{111}E_{222}E_{112}^*E_{221}^*$. With the same method above, it can express as
\begin{eqnarray}\label{equ6}
\fl \langle E_{111} E_{222} E_{112}^{*} E_{221}^{*}\rangle
& =\langle\left(E_{1} T_{1|1} T_{1|1|1}\right)\left(E_{2} T_{2|2} T_{2|2|2}\right)\left(E_{1}^{*} T_{1|1}^{*} T_{2|1| 1}^{*}\right)\left(E_{2}^{*} T_{2|2}^{*} T_{1|2| 2}^{*}\right)\rangle \nonumber \\
& = {\left({\Delta {\omega _1}}\right)^2}\int_{{\omega _0}{\rm{ - }}\frac{1}{2}\Delta {\omega _2}}^{{\omega _0}{\rm{ + }}\frac{1}{2}\Delta {\omega _2}} {{e^{{\rm{ - }}i {{\omega _1}\left( {{t_1} - {t_2}} \right)} }}d} {\omega _1}\int_{{\omega _0}{\rm{ - }}\frac{1}{2}\Delta {\omega _2}}^{{\omega _0}{\rm{ + }}\frac{1}{2}\Delta {\omega _2}} {{e^{ - i{{\omega _2}\left( {{t_2} - {t_1}} \right)} }}d} {\omega _2} \nonumber \\
& = {\left({\Delta {\omega _1}}\right)^2}{\left( {\Delta {\omega _2}} \right)^2}sin{c^2}\left[{\frac{1}{2}\Delta {\omega _2}\left({{t_1} - {t_2}} \right)}\right],
\end{eqnarray}
where $\sin{c}(x)=\sin (x)/x$, the other terms ${E_{112}}{E_{221}}{E_{111}}^*{E_{222}}^*,{E_{121}}{E_{212}}{E_{122}}^*{E_{211}}^*,{E_{122}}{E_{211}}\\\cdot{{E_{121}}^*}{E_{212}}^*$ in the same group have the same result as the one of equation (\ref{equ6}).
All of the terms in the second group add up to $4{\left( {\Delta {\omega _1}} \right)^2}{\left( {\Delta {\omega _2}} \right)^2}sin{c^2}\left[ {\frac{1}{2}\Delta {\omega _2}\left( {{t_1} - {t_2}} \right)} \right]$.

The third term that needs to be calculated is ${E_{111}}{E_{222}}{E_{121}}^*{E_{212}}^*$, it can be obtained by simplification as
\begin{eqnarray}\label{equ7}
\fl \langle E_{111} E_{222} E_{121}^{*} E_{212}^{*}\rangle
& =\langle \left(E_{1} T_{1|1} T_{1|1 | 1}\right)\left(E_{2} T_{2|2} T_{2|2|2}\right)\left(E_{1}^{*} T_{2|1}^{*} T_{1|2| 1}^{*}\right)\left(E_{2}^{*} T_{1|2}^{*} T_{2|1| 2}^{*}\right)\rangle \nonumber \\
& = {\left( {\Delta {\omega _2}} \right)^2}\int_{{\omega _0}{\rm{ - }}\frac{1}{2}\Delta {\omega _1}}^{{\omega _0}{\rm{ + }}\frac{1}{2}\Delta {\omega _1}} {{e^{{\rm{ - }}i {{\omega _{01}}\left( {{t_1}^{''} - {t_2}^{''}} \right)}}}d} {\omega _{01}}\int_{{\omega _0}{\rm{ - }}\frac{1}{2}\Delta {\omega _1}}^{{\omega _0}{\rm{ + }}\frac{1}{2}\Delta {\omega _1}} {{e^{-i{{\omega _{02}}\left( {{t_2}^{''} - {t_1}^{''}} \right)}}}d} {\omega _{02}} \nonumber \\
& = {\left( {\Delta {\omega _1}} \right)^2}{\left( {\Delta {\omega _2}} \right)^2}sin{c^2}\left[ {\frac{1}{2}\Delta {\omega _1}\left( {{t_1}^{''} - {t_2}^{''}} \right)} \right].
\end{eqnarray}
${E_{112}}{E_{221}}{E_{122}}^*{E_{211}}^*$,
${E_{121}}{E_{212}}{E_{111}}^*{E_{222}}^*$,
${E_{122}}{E_{211}}{E_{112}}^*{E_{221}}^*$ are the rest of the terms, they have the same result as equation (\ref{equ7}). The whole third group is $4{\left( {\Delta {\omega _1}} \right)^2}{\left( {\Delta {\omega _2}} \right)^2}sin{c^2}\left[ {\frac{1}{2}\Delta {\omega _1}\left( {{t_1}^{''} - {t_2}^{''}} \right)} \right]$.

The forth term that needs to be calculated is ${E_{111}}{E_{222}}{E_{122}}^ * {E_{211}}^*$ , which is
\begin{eqnarray}\label{equ8}
\fl \langle E_{111} E_{222} E_{122}^{*} E_{211}^{*}\rangle
& =\langle\left(E_{1} T_{1|1} T_{1|1|1}\right)\left(E_{2} T_{2|2} T_{2|2|2}\right)\left(E_{1}^{*} T_{2|1}^{*} T_{2|2| 1}^{*}\right)\left(E_{2}^{*} T_{1|2}^{*} T_{1|1| 2}^{*}\right)\rangle \nonumber \\
& = \int_{{\omega _0}{\rm{ - }}\frac{1}{2}\Delta {\omega _1}}^{{\omega _0}{\rm{ + }}\frac{1}{2}\Delta {\omega _1}} {{e^{{\rm{ - }}i {{\omega _{01}}\left( {{t_1}^{''} - {t_2}^{''}} \right)}}}d} {\omega _{01}}\int_{{\omega _0}{\rm{ - }}\frac{1}{2}\Delta {\omega _2}}^{{\omega _0}{\rm{ + }}\frac{1}{2}\Delta {\omega _2}} {{e^{ - i\left[ {{\omega _{02}}\left( {{t_1} - {t_2}} \right)} \right]}}d} {\omega _{02}} \nonumber \nonumber \\
& \cdot\int_{{\omega _0}{\rm{ - }}\frac{1}{2}\Delta {\omega _1}}^{{\omega _0}{\rm{ + }}\frac{1}{2}\Delta {\omega _1}} {{e^{{\rm{ - }}i{{\omega _{01}}\left( {{t_2}^{''} - {t_1}^{''}} \right)}}}d} {\omega _{01}}\int_{{\omega _0}{\rm{ - }}\frac{1}{2}\Delta {\omega _2}}^{{\omega _0}{\rm{ + }}\frac{1}{2}\Delta {\omega _2}} {{e^{ - i\left[ {{\omega _{02}}\left( {{t_2} - {t_1}} \right)} \right]}}d} {\omega _{02}} \nonumber \\
& = {\left( {\Delta {\omega _1}} \right)^2}{\left( {\Delta {\omega _2}} \right)^2}sin{c^2}\left[ {\frac{1}{2}\Delta {\omega _1}\left( {{t_1}^{''} - {t_2}^{''}} \right)} \right]sin{c^2}\left[ {\frac{1}{2}\Delta {\omega _2}\left( {{t_1} - {t_2}} \right)} \right].
\end{eqnarray}
And the rest of the terms,
${E_{112}}{E_{221}}{E_{121}}^*{E_{212}}^*$,
${E_{121}}{E_{212}}{E_{112}}^*{E_{221}}^*$,
${E_{122}}{E_{211}}{E_{111}}^*{E_{222}}^*$ in the same group have the same result as the one of equation (\ref{equ8}). We're going to add up to $4{\left( {\Delta {\omega _1}} \right)^2}{\left( {\Delta {\omega _2}} \right)^2}sin{c^2}\left[ {\frac{1}{2}\Delta {\omega _1}\left( {{t_1}^{''} - {t_2}^{''}} \right)} \right]sin{c^2}\left[ {\frac{1}{2}\Delta {\omega _2}\left( {{t_1} - {t_2}} \right)} \right]$.

Finally, when all the above terms are added together, the second-order temporal correlation function with two RGs in the scheme as shown in figure \ref{figure1} is
\begin{eqnarray}\label{equ9}
\fl {G^{(2)}}({t_1} - {t_2};{t_1}^{''} - {t_2}^{''})&=\frac{1}{4}{\left( {\Delta {\omega _1}} {\Delta {\omega _2}} \right)^2}({1 + sin{c^2}\left[ {\frac{1}{2}\Delta {\omega _1}\left({{t_1}-{t_2}} \right)} \right] +sin{c^2}\left[{\frac{1}{2}\Delta {\omega _2}\left( {{t_1}-{t_2}} \right)} \right]} \nonumber \\
&+ sin{c^2}\left[ {\frac{1}{2}\Delta {\omega _1}\left( {{t_1}^{''}-{t_2}^{''}} \right)} \right]sin{c^2}\left[ {\frac{1}{2}\Delta {\omega _2}\left( {{t_1} - {t_2}} \right)} \right]),
\end{eqnarray}
where ${\tau _1} = {t_1}^{''} - {t_2}^{''}$, ${\tau _{\rm{2}}} = {t_1} - {t_2}$, they represent the coherent time that the laser passing through the first and second RG, respectively. Taking these equations into equation (\ref{equ9}), the normalized
second-order temporal correlation function can be expressed as
\begin{equation}\label{equ10}
{g^{(2)}}({\tau _1};{\tau _{\rm{2}}}) = \left[ {1 + sin{c^2}\left( {\frac{1}{2}\Delta {\omega _1}{\tau _1}} \right)} \right]\cdot\left[ {1 + sin{c^2}\left( {\frac{1}{2}\Delta {\omega _{\rm{2}}}{\tau _2}} \right)} \right].
\end{equation}

When the value of ${\tau _1}$ and ${\tau _2}$ approaches infinity,  ${g^{(2)}}({\tau _1};{\tau _{\rm{2}}})$ equals 1, which means detection events are independent each other. When ${\tau _1}$  and ${\tau _2}$ all equals zero, ${g^{(2)}}({\tau _1};{\tau _{\rm{2}}})$ equals 4, which means superbunching effect can be observed. Also the result of theoretical derivation is in agreement with the conclusion derived from the quantum two-photon interference theory \cite{zhou2017superbunching}.

\subsection{Experimental verification}
\begin{figure}[htb]
\centering\includegraphics[width=12cm]{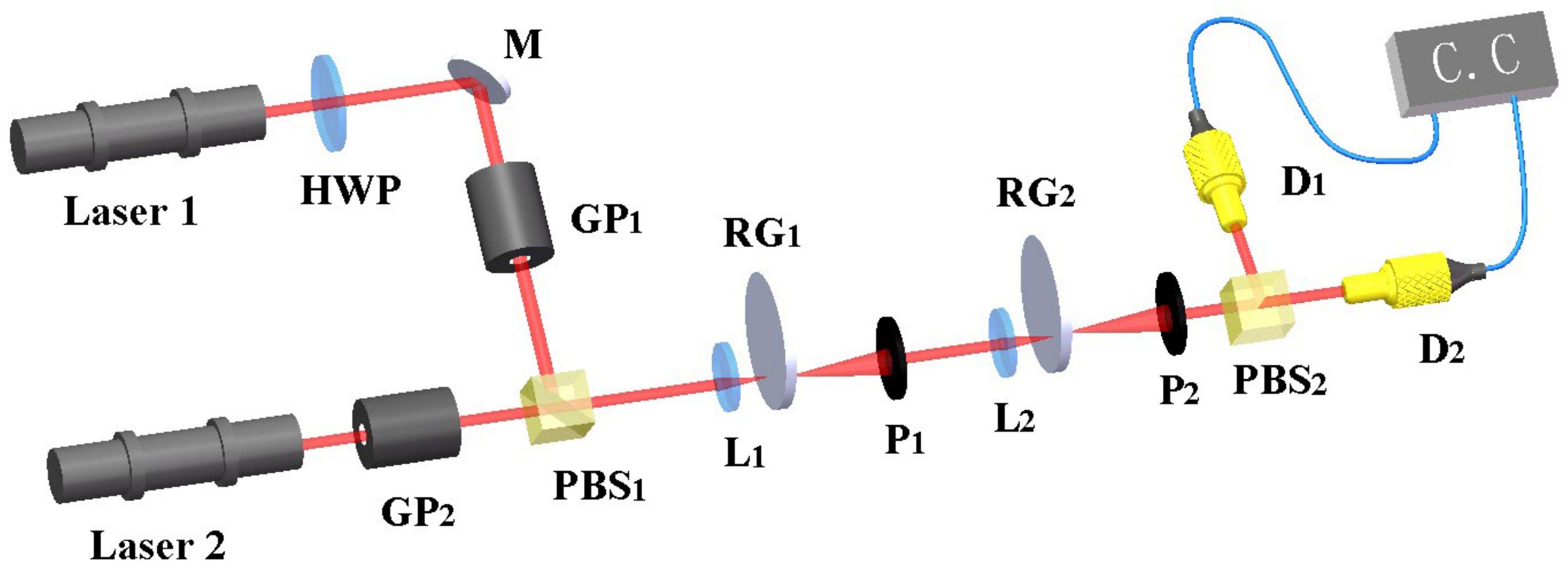}
\caption{\label{figure2} Experimental setup for measuring superbunching effect from a pair of modulated distinguishable classical light. Laser 1 and Laser 2 are single-mode continuous-wave He-Ne lasers. $\rm PBS_1$ and $\rm PBS_2$ are two polarized beam splitters. $\rm RG_1$ and $\rm RG_2$ are two rotating ground glass plates that can adjust the rotational speed. $\rm P_1$ and $\rm P_2$ are the pinholes. The measuring system is a HBT-like intensity interferometer.}
\end{figure}

The experimental setup for observing two-photon superbunching from a pair of modulated distinguishable classical light is shown in figure \ref{figure2}. The initial polarization of two employed He-Ne lasers are horizontal. A half-wave plate (HWP) behind laser 1 is employed to change the horizontal polarization to vertical polarization. $\rm GP_1$ and $\rm GP_2$ are two Glan prisms, which are set to purify the polarization of these two lasers. $\rm L_1$ and $\rm L_2$ are convergent lenses with focal lengths of 50 mm. The distance between $\rm RG_1$ and $\rm P_1$ is 300 mm. The transverse coherence length of pseudothermal light generated by $\rm RG_1$ is 1.9 mm in the plane of $\rm P_1$. The diameter of the $\rm P_1$ is 1 mm, which is less than the coherence length in order to pass through $\rm P_1$ within one same coherence area. $\rm L_2$ is also employed to focus the light onto $\rm RG_2$. The distance between $\rm L_2$ and $\rm RG_2$ needs to be larger than the focus length of $\rm L_2$ mainly because pseudothermal light scattered by $\rm RG_1$ is not parallel. $\rm D_1$ and $\rm D_2$ are two single-photon detectors (PerkinElmer, SPCM-AQRH-14-FC). C.C is a two-photon coincidence count detection system (Becker \&. Hickl GmbH, DPC-230). The experiment mainly includes the following three steps.

\begin{figure}
\centering\includegraphics[width=12cm]{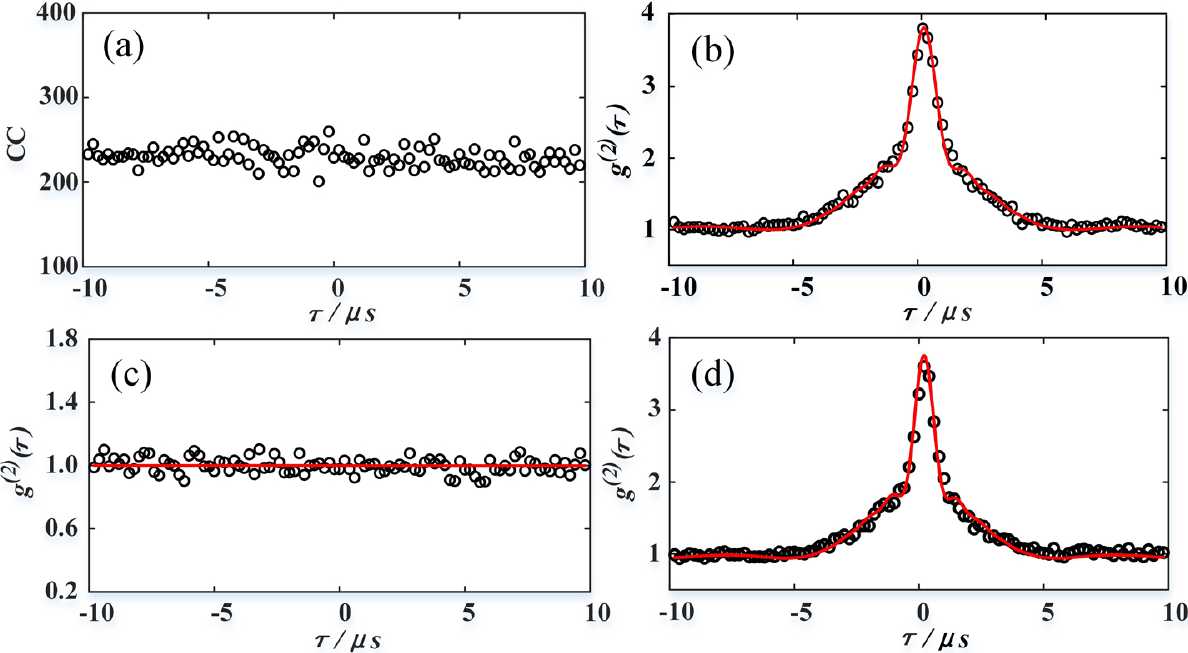}
\caption{\label{figure3} (a) Two-photon coincidence
counts when only turn on laser 1. (b) The degree of second-order temporal correlation when only $45^{\circ}$ linearly polarized light comes in. (c) The result of measured normalized second-order temporal correlation functions when two pseudothermal lights focus on different areas of $\rm RG_1$. (d) The result of measured normalized second-order temporal correlation functions when two pseudothermal lights focus on same areas of $\rm RG_1$. $\tau$ is the time difference between two single-photon detection
events within a two-photon coincidence count. The black circles are measured results and the red lines are theoretical fittings.}
\end{figure}

In the first step, turn on laser 1 and turn off laser 2, let the vertically polarized light pass through $\rm RG_1$ and $\rm RG_2$ successively. The traditional second-order temporal correlation function of the ordinary pseudothermal superbunching effect is measured to verify the reliability of the experimental test system. $\rm RG_1$ and $\rm RG_2$ are rotating at 100 Hz and 50 Hz, respectively. As a result of the vertical polarization, the modulated light passes through the $\rm PBS_2$ and only arrives at $\rm D_1$. The experimental results are shown in figure \ref{figure3}(a). The coincidence count is almost constant for 50s of collection time. It has no superbunching effect obviously because there is almost no light reaching $\rm D_2$. Keeping experimental devices unchanged, when we rotate additional HWP to make it $45^{\circ}$ polarized light respect to the horizontal direction which is placed between the $\rm PBS_1$ and the $\rm L_1$ (not shown in figure \ref{figure2}), the pseudothermal light will be split into two beams with the same intensity after passing through the $\rm PBS_2$. Finally, the HBT intensity interferometer will be triggered by two pseudothermal lights. The results are shown in figure \ref{figure3}(b). The degree of second-order correlation ${g^{\left(2\right)}}\left(0\right) = 3.91\pm0.02$ was observed, which means the superbunching effect of pseudothermal light was achieved. Also the measured full width at half maximum (FWHM) reaches about $1.7\mu s\pm 0.02$ and the visibility of the peak is about 59.2\%.

In the second step, turn on the laser 1 and laser 2 simultaneously. Vertically polarized light from laser 1 and horizontally polarized light from laser 2 are combined into one beam at the $\rm PBS_1$, and then focus on $\rm RG_1$ by the lens $\rm L_1$. Note that two beams focus on different areas of $\rm RG_1$ at the moment. That means $\rm RG_1$ will produce two completely different sets of speckles. When two sets of speckles with mutually perpendicular polarization directions pass through $\rm PBS_2$, the vertically polarized light from the laser 1 will reach the detector $\rm D_1$, and the horizontally polarized light from laser 2 will reach the detector $\rm D_2$. The measurement result is shown in figure \ref{figure3}(c), ${g^{\left(2\right)}}\left(\tau\right)$ is flat and no superbunching effect occurs.

In the third step, except for the condition that the combination of the two light beams from laser 1 and laser 2 through $\rm PBS_1$ focus on the same area of $\rm RG_1$, the rest of experimental steps are the same as the second steps. Hence the speckle patterns scattered by $\rm RG_1$ and $\rm RG_2$ are almost the same, and they have the same temporal and spatial fluctuations. The measurement result is shown in figure \ref{figure3}(d). The circles is the measured result, the red line is theoretical fittings by employing equation (\ref{equ9}). The FWHM is $1.71 \mu s \pm 0.02$ and the visibility of the peak is about 58.5\%. The degree of second-order correlation is
${g^{\left(2\right)}}\left(0\right)=3.83\pm0.02$ as depicted in the figure \ref{figure3}(d). It is obviously that ${g^{\left(2\right)}}\left(0\right)$ is greater than 2, which means the temporal superbunching effect of pseudothermal light was observed successfully.

\begin{figure}[htb]
\centering\includegraphics[scale=0.3]{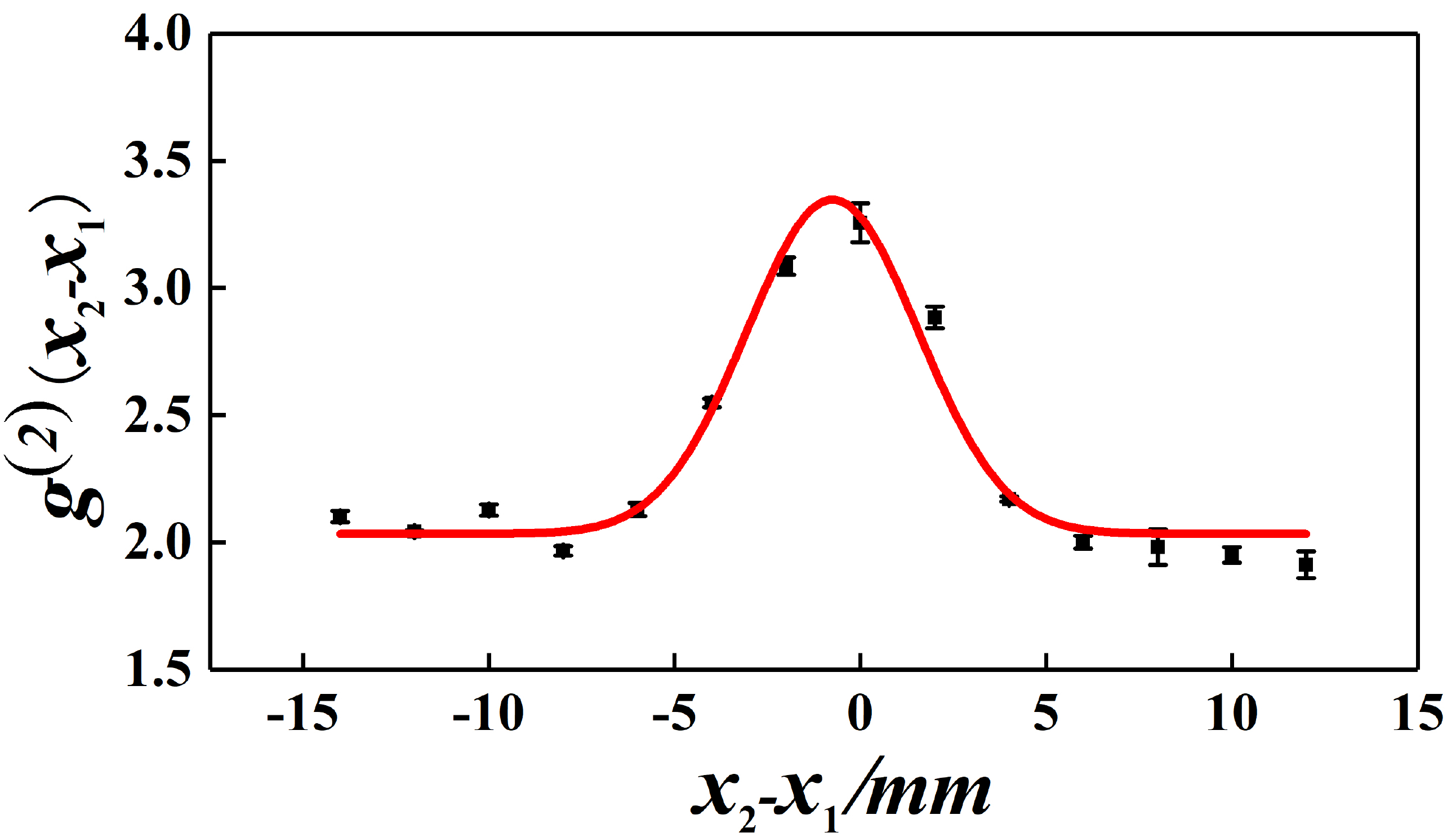}
\caption{\label{figure4} The result of measured normalized second-order spatial correlation functions. The black circles are measured results and the red line is theoretical fitting.}
\end{figure}

Based on the the third step, we proceed to study the superbunching effect in the spatial domain. With $\rm D_1$ fixed at $x_1=0$, $\rm D_2$ is then moved transversely in steps of 2 mm through $x_2= \pm12 mm$. The collection time for every steps is 50s. The spatial normalized second-order correlation function ${g^{\left(2\right)}}\left(x_1-x_2\right)$ is calculated and the second-order correlation pattern as shown in figure \ref{figure4}. The FWHM of this peak is $4.8 \pm 0.01 mm$, while the visibility of the peak is about 22.4\%. The degree of second-order correlation of pseudothermal light was measured as ${g^{\left(2\right)}}\left(0\right)=3.42\pm0.02$, which means superbunching effect in the spatial domain was also observed.

\section{\label{Discussion}Discussion}
The superbunching effect was observed in the first step. From a classical theory of view, when the $45^{\circ}$ linearly polarized light passes through $\rm RG_1$, $\rm RG_2$ and $\rm PBS_2$ in sequence, two single photon detectors at symmetric position are triggered by the pseudothermal light which possesses the same intensity fluctuation and photon statistical distribution. Therefore, the classical intensity fluctuations theory \cite{goodman2007speckle,ou1988violation} could be explained why the superbunching effect can be observed. In the quantum two-photon interference theory \cite{mandel1999quantum}, when we changed the horizontal light to $45^{\circ}$ linearly polarized light by rotating the HWP, the pseudothermal lights pass through $\rm PBS_2$, there are two different and indistinguishable paths to trigger two detectors. It is the reason why we can use quantum two-photon interference theory to explain the superbunching effects, simultaneously the superbunching effects of ${g^{\left(2\right)}}\left(0\right) = 3.91\pm0.02$ was observed.

On the basis of the first step, turn on the laser 1 and laser 2 simultaneously, the second step and third step are a group of comparative experiments. The difference between them is whether two lights focus on same areas of $\rm RG_1$ or not. From the classical theory of intensity fluctuation correlation, the second step and the third step will get different results. When the lasers focus on the $\rm RG_1$ in different positions, it means the two-photon coincidence count system is triggered by two completely different sets of speckles. Therefore, the superbunching effects can not be observed corresponding to figure \ref{figure3}(c). The third step is the opposite of the second step, the superbunching effects is observed corresponding to figure \ref{figure3}(d) mainly because the lasers focus on the $\rm RG_1$ in same area. The two set of speckles that pass through $\rm RG_1$ and $\rm RG_2$ own identical fluctuations in the same coherent region.

However, the quantum two-photon interference theory will give same prediction results between the second step and the third step. It is well known that two-photon interference theory emphasizes whether the interference path is different and indistinguishable. In the second and third step, the polarization directions of two incident lasers are all orthogonal. After passing through $\rm PBS_2$, the horizontally polarized light enters the detector $\rm D_2$, and the vertically polarized light enters the detector $\rm D_1$. There is only one path to trigger the two photon coincidence count detection system. Under the prediction of quantum two-photon interference theory, there is no superbunching effect in the second and third step. But we observed the superbunching effect of pseudothermal light in the third step, which the degree of second-order temporal correlation was measured as ${g^{\left(2\right)}}\left(0\right)=3.83\pm0.02$ as shown in figure \ref{figure3}(d). It is contrary to the prediction of classical theory of intensity fluctuation correlation. The temporal superbunching effect of classical light in the figure \ref{figure3}(d) seems to break the unity of classical and quantum theory. The quantum two-photon interference theory can not explain this strange phenomenon. Here we used the the classical intensity fluctuations theory to explain the superbunching effect without two-photon interference.

On account of that the polarization of two beams are orthogonal and they focus on the same area of $\rm RG_1$, two pseudothermal light beams own the different polarization direction but generate the same fluctuation. It means the distribution of electric field ${E}({r},{t})$ is the same. After transferring through $\rm RG_1$ and $\rm RG_2$ in series, the two-photon coincidence count system is triggered by two sets of same speckles. The specific derivation is shown in the above Sec. \ref{Methods}. Therefore, the superbunching effect can be well explained by the classical intensity fluctuations theory.

In the fourth step, the background is equal to 2 in the spatial superbunching correlation diagram. This is mainly due to the fact that $\rm P_1$, which is set in front of $\rm RG_2$, performs a spatial mode selection on the light passing through $\rm RG_2$ so that the speckle has the same fluctuation in the same spatial mode. However, there is no change in the longitudinal correlation length related to the superbunching effect in the time domain. Therefore, the temporal superbunching effects is different from the spatial superbunching effect, and the measured background of the superbunching effect is 1 and 2, respectively.

It is well known that there exists space-time duality between the equations that describe the paraxial diffraction of light beams \cite{Kolner:89,301659,TORRESCOMPANY20111}. Similar to the above methods for calculating the temporal superbunching effect, the classical intensity fluctuation correlation theory can also be used to calculate the spatial superbunching effect. The expression of electromagnetic wave which transmitted to the surface of detector can be expressed as
\begin{eqnarray}\label{equ11}
E_{i j k} &=E_{i} \cdot T_{j | i} \cdot T_{k|j|i} \nonumber \\
& =e^{-ik_{0}\left(r_{i}-r_{0}\right)} \int_{-\frac{1}{2} d_{1}}^{\frac{1}{2}d_{1}} e^{-ik_{0j}\left(r_{j}^{''}-r_{i}^{'}\right)} d r_{0 j} \int_{-\frac{1}{2}d_{2}}^{\frac{1}{2} d_{2}} e^{-ik_{k}\left(r_{k}-r_{j}^{''}\right)} d r_{k}.
\end{eqnarray}

The meaning of the symbol in the above equation (\ref{equ11}) is similar to equation (\ref{equ2}). $d_1$ and $d_2$ represent the length of the pseudothermal light source after passing through $\rm RG_1$ and $\rm RG_2$, respectively. ${k_0}$ is the wave vector of the laser, ${k _{0i}}$ is the wave vector of the light scattered by the position $i$ of the $\rm RG_1$, and ${k _{j}}$ is the wave vector of the light scattered by the position of the $\rm RG_2$. $r_i^{'}$, $r_j^{''}$ and $r_k$ are position vectors when the electromagnetic wave arrives at position $i$ of the $\rm RG_1$, position $j$ of the $\rm RG_2$ and the surface $k$ of detector respectively, where $i,j,k = 1,2$. Substituting the equation (\ref{equ11}) into equation (\ref{equ4}) and the normalized second-order temporal correlation function is
\begin{equation}\label{equ12}
{g^{(2)}}(\Delta {x_1};\Delta {x_2}) = \left[ {1 + sin{c^2}\left( {\frac{{\pi L_1}}{{\lambda d_1}}\Delta {x_1}} \right)} \right]\cdot\left[{1 + sin{c^2}\left({\frac{{\pi L_2}}{{\lambda d_2}}\Delta {x_2}} \right)} \right],
\end{equation}
we define $\Delta {x_1} = x_1^{'} - x_2^{'}$ that means the position difference between two orthogonal beams of light propagating over $\rm RG_1$, $\Delta {x_2} = x_1 - x_2$ means the transverse position difference between detectors $\rm D_1$ and $\rm D_2$. $\lambda$ is the wavelength of the light source, $\rm L_1$ is the distance from $\rm RG_1$ to detectors and $\rm L_2$ is the distance from $\rm RG_2$ to detectors.

As a result of two orthogonal light beams focus on the same area of $\rm RG_1$ ($\Delta {x_1} = 0$), they generate same fluctuation after passing through $\rm RG_1$. So the equation (\ref{equ12}) can be further simplified as
\begin{equation}\label{equ13}
{g^{(2)}}(\Delta {x_2})=2\left[{1+sin{c^2}\left({{\frac{{\pi L_2}}{{\lambda d_2}}}\Delta {x_2}}\right)}\right],
\end{equation}
when the value of $\Delta {x_2}$ is infinity, ${g^{(2)}}(\Delta {x_2})$ equals 2, which means the two detectors are far enough apart; ${g^{(2)}}(\Delta {x_2})$ equals 4 when $\Delta {x_2}=0$, which means two detectors are in the same symmetric position.

The background is close to 2 in our experiment about spatial superbunching effect. The reasons are as follows, taking off $\rm RG_2$ and repeating the third step above, the measured degree of second-order temporal correlation is $1.90 \pm 0.01$. It is the bunching effect that ${g^{\left(2\right)}}\left(0\right)=2$. When we add the $\rm RG_2$, the modulated light is modulated again. This is the reason why we can observe spatial superbunching effect. At the same time, the vibration of $\rm RG_1$ and $\rm RG_2$ would cause the vibration of the optical table in the actual experiment, so the measurement of the degree of second-order correlation is 1.9 or higher because of the vibration of laser. Therefore, the measured background of spatial superbunching effect accords with our theory that equals 2.

\section{\label{Conclusion}Conclusion}
In summary, we achieved the superbunching effect from a pair of modulated distinguishable classical light in the temporal and spatial domain. We employed the classical theory of intensity fluctuation correlation to deduce the normalized second-order correlation function of temporal and spatial superbunching effect. It shows good agreement with the measurement as ${g^{\left(2\right)}}\left(0\right)=3.83\pm0.02$. Also the degree of second-order spatial correlation was measured as $3.42\pm0.02$. Although there are completely opposite predictions by using the classical theory of intensity fluctuation correlation and quantum two-photon interference theory to explain the phenomena in the experiment, it does not mean quantum theory contradicts classical theory. It only shows there is no two-photon interference phenomenon in this experiment. Therefore, studying this interesting phenomenon is not only conducive to the future research about superbunching effect that have potential application in improving the visibility of ghost imaging, but also plays an important role in understanding the relationship between classical theory of intensity fluctuation correlation and quantum two-photon interference theory.

\section*{Acknowledgments}
The authors would like to thank Dr. J.B. Liu for helpful discussions. This work was supported by Shaanxi Key Research and Development Project (Grant No. 2019ZDLGY09-10); Key Innovation Team of Shaanxi Province (Grant No. 2018TD-024) and 111 Project of China (Grant No.B14040).

\section*{ORCID iDs}
Sheng Luo https://orcid.org/0000-0001-6495-0900\\
Huai-Bin Zheng https://orcid.org/0000-0003-2313-4119

\bibliographystyle{iopart-num}
\bibliography{reference}

\end{document}